\documentclass[conference]{IEEEtran}
\IEEEoverridecommandlockouts
\usepackage{cite}
\usepackage{amsmath,amssymb,amsfonts}
\usepackage{algorithmic}
\usepackage{graphicx}
\usepackage{textcomp}
\usepackage{xcolor}
\usepackage{cite}
\usepackage{longtable}
\usepackage{multirow} 
\usepackage[numbers,sort&compress]{natbib}
\usepackage[T1]{fontenc}
\usepackage{caption}
\usepackage{listings}
\usepackage{upquote}
\usepackage{xcolor}
\usepackage[normalem]{ulem}
\begin{document}

\title{Enhancing Public Speaking Skills in Engineering Students Through AI\\

}
\author{
  \IEEEauthorblockN{
    Amol Harsh\IEEEauthorrefmark{1},
    Brainerd Prince\IEEEauthorrefmark{2},
    Siddharth Siddharth\IEEEauthorrefmark{1},
    Deepan Raj Prabakar Muthirayan\IEEEauthorrefmark{1},\\
    Kabir S Bhalla\IEEEauthorrefmark{1},
    Esraaj Sarkar Gupta\IEEEauthorrefmark{1},
    Siddharth Sahu\IEEEauthorrefmark{1}
  }
  \\
  \IEEEauthorblockA{\IEEEauthorrefmark{1}
    Computer Science and Artificial Intelligence\\
    Plaksha University, Mohali, India\\
    Email: \{amol.harsh, siddharth.s, deepan.muthirayan,\\
    kabir.bhalla.ug24, Esraaj.Gupta.ug24, Siddharth.sahu\}@plaksha.edu.in
  }
  \\
  \IEEEauthorblockA{\IEEEauthorrefmark{2}
    Center for Thinking, Language and Communication\\
    Plaksha University, Mohali, India\\
    Email: brainerd.prince@plaksha.edu.in
  }
}

\maketitle

\begin{abstract} 
This research-to-practice full paper was inspired by the persistent challenge in effective communication among engineering students. Public speaking is a necessary skill for
future engineers as they have to communicate technical knowledge with diverse stakeholders. While universities offer courses or workshops, they are unable to offer sustained and personalized training to students. Providing comprehensive feedback on both verbal and non-verbal aspects of public speaking is time-intensive, making consistent and individualized assessment impractical. This study integrates research on verbal and non-verbal cues in public
speaking to develop an AI-driven assessment model for engineering students. Our approach combines speech analysis, computer vision, and sentiment detection into a multi-modal AI system that provides assessment and feedback. The model evaluates (1) verbal
communication (pitch, loudness, pacing, intonation), (2) non-
verbal communication (facial expressions, gestures, posture), and
(3) expressive coherence, a novel integration ensuring alignment
between speech and body language. Unlike previous systems
that assess these aspects separately, our model fuses multiple
modalities to deliver personalized, scalable feedback. Preliminary testing demonstrated that our AI-generated feedback was moderately aligned with expert evaluations. Among the state-of-the-art AI models evaluated—all of which were Large Language Models (LLMs), including Gemini and OpenAI models—Gemini Pro emerged as the best-performing, showing the strongest agreement with human annotators. 
By eliminating reliance on human evaluators, this AI-driven public speaking trainer enables repeated practice, helping students naturally align their speech with body language and emotion—crucial for impactful and professional communication.

\end{abstract}


\section{Introduction}

Effective communication is an essential skill for engineers, critical for bridging the gap between technical innovation and societal impact. Engineering professionals regularly interact with diverse stakeholders, many of whom lack technical expertise. Consequently, the ability to clearly and effectively convey complex ideas is not only beneficial but imperative for successful project implementation, stakeholder engagement, and career advancement. Effective communication facilitates collaboration, innovation dissemination, and informed decision-making, directly contributing to organizational and societal progress \cite{sageev2001message}.

Despite its importance, current engineering curricula at universities frequently underrepresent comprehensive communication training, limiting exposure mostly to isolated workshops or occasional project presentations. This restricted approach is insufficient to cultivate the nuanced and sustained development required for adept public speaking \cite{pitt2011publicspeaking}. Moreover, effective communication encompasses both verbal and non-verbal aspects—ranging from vocal modulation and clarity to gestures and facial expressions \cite{wu2023evaluating}. Traditional methods of communication training often fail to integrate these dimensions effectively, leaving a gap in students’ ability to master and harmonize their verbal and non-verbal communication skills.

Another significant limitation arises from the reliance on human evaluators for providing individualized feedback on public speaking skills. Human evaluation, while effective, is inherently resource-intensive and subject to variability. The scalability of personalized feedback becomes problematic, especially in large classes typical of engineering institutions. Students thus face inconsistent or infrequent feedback, which impedes their ability to systematically improve their public speaking capabilities \cite{liow2012assessment}.

To address these challenges, this paper proposes a novel multi-modal Large Language Model (LLM)-based evaluator designed specifically for assessing and enhancing public speaking skills among engineering students. This advanced AI-driven approach seamlessly integrates speech analysis, computer vision, and sentiment detection to assess verbal parameters such as pitch, pacing, and intonation, alongside non-verbal elements including facial expressions, posture, and gestures. Finally, a novel contribution of this research is the introduction of a new concept, \textit{expressive coherence}, an innovative measure ensuring alignment between verbal articulation and corresponding non-verbal cues, thereby offering a holistic assessment of communication effectiveness. Our literature review has revealed that this aspect of public speaking has not been explicitly captured in public speaking assessments.

The broader implications of this research extend significantly beyond educational contexts:

\begin{enumerate}
    \item \textbf{Transforming Scalable Communication Training Across Disciplines.}
    Although designed for engineering students, this multi-modal AI framework can extend to broader educational and professional contexts. By automating feedback, it reduces reliance on one-on-one coaching and scales personalized training to larger cohorts, potentially reshaping how institutions teach and assess oral communication.

    \item \textbf{Pioneering a More Holistic Metric for Human-AI Interaction.}
    The concept of “expressive coherence” unites verbal and non-verbal cues, emphasizing their synergistic role in effective speaking. This metric paves the way for AI-driven evaluations that capture how speech, gestures, and emotional signals align—offering applications in remote collaboration, mental health assessments, and other human–AI interaction fields.
\end{enumerate}

Through this innovative multi-modal assessment, our research aims to transform public speaking training, enhancing both academic and professional communication standards.

\section{Literature Survey}
\maketitle

\begin{figure}[t]
\centerline{\includegraphics[width=\linewidth]{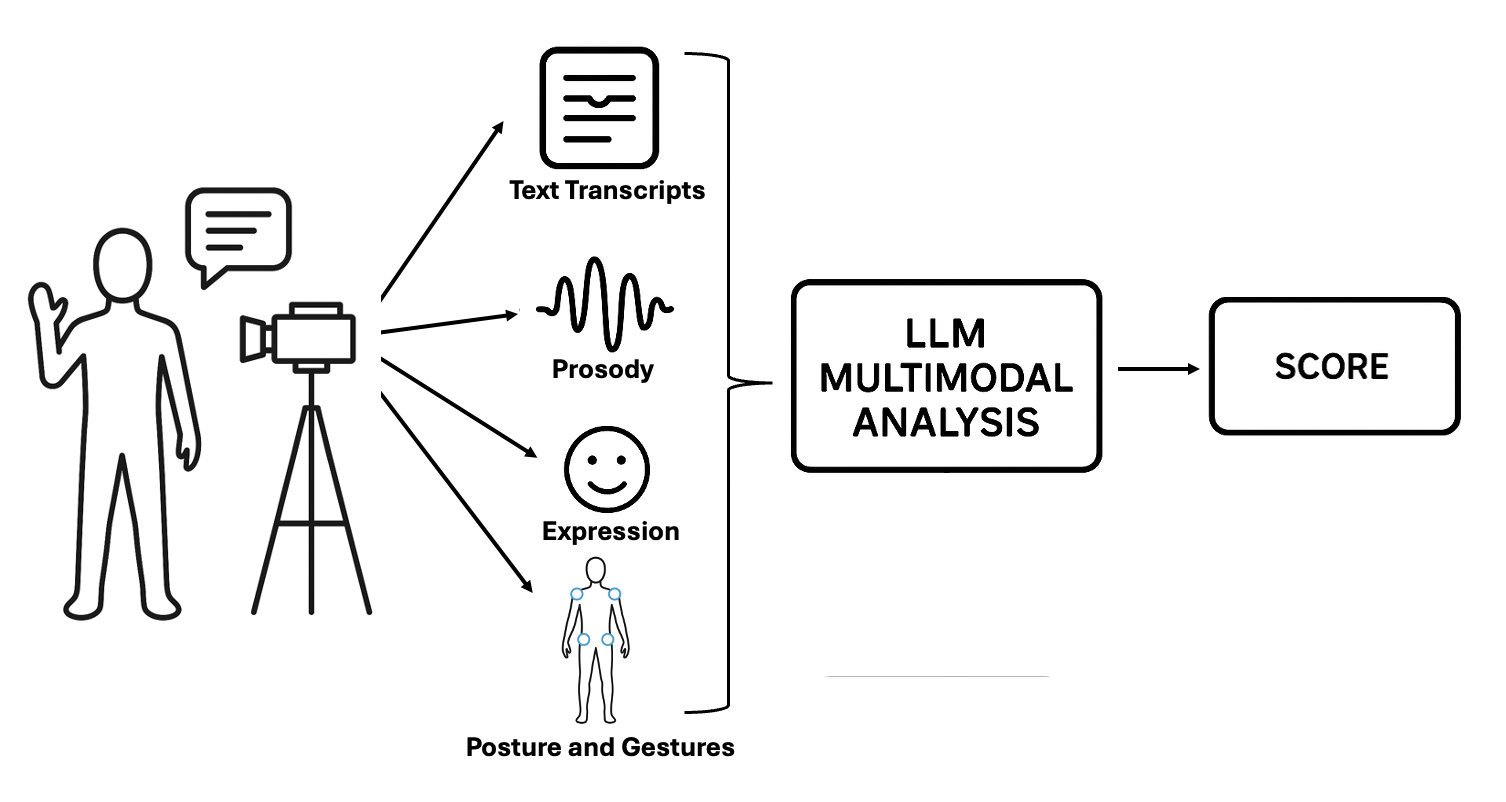}}
\caption{Overview of the AI-powered public speaking evaluation pipeline.}
\label{fig:ai_pipeline}
\end{figure}

\begin{figure*}[ht]
\centering
\begin{minipage}[c]{0.9\textwidth}
\begin{lstlisting}[
    language=Python,
    caption={An example snippet of rich multi-modal data combining text, vocal, and non-verbal features.},
    label={lst:multi-modal_snippet},
    basicstyle=\footnotesize\ttfamily,
    breaklines=true,
    frame=single,
    numbers=left,
    numberstyle=\tiny,
    xleftmargin=2em,
    framexleftmargin=1.5em
]
[70.0 - 80.0 "secs"]: {
 "transcript": "Now, we try to evaluate the trade off inaccuracy and energy consumption if we were to use these SLMs in scale instead of LLMs.",
  "posture": ["Upright"],
  "pose": ["Open Pose (Arms Uncrossed)"],
  "pitch": ["Normal", "High", "Low", "Normal", "Low", "Normal", "Normal", "Normal", "Low", "Low"],
  "loudness": ["Normal", "High", "Normal", "Normal", "High", "Normal", "Normal", "Low", "High", "Low"],
  "speech_rate": "144.00 words per minute",
  "intonation_pattern": ["Normal", "High", "Normal", "High", "High", "High", "High", "Normal", "Normal", "Low"],
  "face_expression": ["neutral"],
  "horizontal_gesture": ["high unilateral gesture", "medium wide unified gesture towards the left", "high wide unified gesture towards the right"
  ],
  "vertical_gesture": [ "high area unified falling gesture","normal area unified falling gesture","normal unilateral down or up gesture", "normal unilateral down or up gesture", "high area unified rising gesture"
  ],
  "hand_configuration": ["Left hand: open, Right hand: open", "Hands on top of each other: No"
  ]
}
\end{lstlisting}
\caption{An example snippet of rich multi-modal data combining text, vocal, and non-verbal features.}
\label{fig:multi-modal_snippet}
\end{minipage}
\end{figure*}
Engineers’ career success and educational outcomes are closely linked to communication skills, especially oral communication and public speaking. These abilities are seen as critical for employability, interdisciplinary collaboration, and career advancement. Studies highlight their value across alumni surveys, workplace interviews, and employer assessments, often ranking them as highly as technical competencies. Engineering education is thus increasingly urged to integrate structured public speaking training \cite{sageev2001message, passow2012abet, darling2003importance, wu2023evaluating, coffelt2024communication}.

Despite its importance, teaching and assessing public speaking in engineering programs face significant challenges. Traditional academic settings rely heavily on in-class presentations evaluated by instructors or peers, which can be subjective, time-consuming, and difficult to scale for large groups. Research has highlighted that most public-speaking assessments in educational contexts depend on human evaluators, which limits opportunities for students to receive repeated feedback \cite{chen2016automated}. The time-intensive nature of assessment and feedback has also been identified as a major hurdle in the development of oral presentation skills in engineering programs \cite{degrez2010response}. One approach to overcome the constraint of limited assessment time was the incorporation of peer review. However, discrepancies between peer and instructor scoring introduce subjectivity and inconsistency, thereby limiting the effectiveness of peer review as a form of assessments \cite{liow2012assessment}. These constraints—limited time, heavy reliance on human raters, and lack of consistent feedback—mean that many engineering students graduate without having fully developed or tested their public speaking skills.

While artificial intelligence has been explored as a solution to the shortcomings of traditional public speaking training, most existing systems tend to concentrate on isolated surface-level delivery features—such as vocal tone, posture, or gesture—without capturing the cognitive or semantic depth of an effective presentation \cite{kurihara2007presentation, batrinca2013cicero}. Even systems that integrate multi-modal feedback, like Cicero and Presentation Sensei, typically evaluate modalities independently and fail to assess how well verbal and non-verbal cues work together to convey meaning \cite{kurihara2007presentation, batrinca2013cicero, tanveer2015rhema, schneider2015presentation}. Furthermore, most of these tools do not attempt to model the alignment between what is said and how it is expressed emotionally and physically. While recent advances—such as the work by \cite{barkar2025decoding}—have begun testing LLMs to evaluate aspects like persuasiveness, their focus remains limited to textual content alone. No prior study, to our knowledge, has introduced a cohesive evaluative framework that integrates verbal content, non-verbal delivery, and emotional alignment into a unified metric. In contrast to systems that treat each modality separately, we aim to introduce a new metric—\textit{Expressive Coherence}—which evaluates how dynamically a speaker emphasizes key points and aligns emotional tone through both speech and behavior. 

To address the limitations of existing AI systems that evaluate verbal and non-verbal cues in isolation, we propose, SapienAI, a unified LLM-based evaluation framework that analyzes all data modalities—verbal, non-verbal, and their interplay—simultaneously during grading (Figure \ref{fig:ai_pipeline}). Unlike earlier approaches that treat each modality as a separate scoring dimension \cite{beckner2024unlocking, padia2024ai}, our system enables the model to process and interpret multi-modal cues holistically, offering a richer and more context-aware understanding of student presentations. By integrating speech transcripts, vocal dynamics, facial expressions, and body language into a single LLM-driven pipeline, the model can evaluate not just what is being said and how it is delivered, but also how well these elements align with each other. This fusion of modalities, supported by the interpretive capabilities of LLMs, allows us to provide more nuanced, scalable, and pedagogically meaningful assessment for public speaking training.

\section{Methodology}

\subsection{Data Collection}

A total of 20 undergraduate students (age range 18–20) from a technology-focused university volunteered for this study. Each participant provided informed consent prior to data collection and was briefed on the study’s objectives. Ethical guidelines were strictly followed, ensuring that participants’ privacy and confidentiality were maintained throughout the research process.

Each participant was instructed to prepare and then deliver a brief (approximately two-minute) presentation explaining a technical concept to a non-technical audience. The presentations took place in a controlled setting equipped with a high-definition video camera and a dedicated microphone to ensure clear visual and audio capture. Four audience members were present in the room to replicate a realistic speaking environment; these audience members were non-technical individuals, thereby reinforcing the challenge of tailoring technical content in an accessible manner.

During each presentation, participants were free to use any speaking style or gesture repertoire that felt natural to them. No specific guidelines on posture, gesturing, or vocal delivery were provided, so that the data would genuinely reflect each speaker’s natural public-speaking tendencies.

\begin{table*}[ht!]
\captionsetup[table]{skip=10pt}
\small
\renewcommand{\arraystretch}{1.2}
\setlength{\tabcolsep}{4pt}
\centering
\begin{tabular}{|p{3cm}|p{3cm}|p{5cm}|p{5cm}|}
\hline
\textbf{Category} & \textbf{Sensing Modality} & \textbf{Physiological Response} & \textbf{Response Indicator} \\
\hline
\multirow{2}{*}{Body Posture/Pose} 
    & Posture 
    & Upright / Not Upright 
    & Confidence, attentiveness / Lower confidence, inattentiveness \\ \cline{2-4}
    & Pose 
    & Open (Arms Uncrossed) / Closed (Arms Crossed) 
    & Indicates receptiveness, engagement / Signals defensiveness or reservation \\ \hline
\multirow{4}{*}{Voice Prosody} 
    & Pitch 
    & Low / Normal / High 
    & Deep, subdued (calm/serious) / Standard tone / Elevated (excited/emphatic) \\ \cline{2-4}
    & Loudness (RMS Energy) 
    & Low / Normal / High 
    & Soft, intimate, hesitant / Typical volume / Forceful, urgent, intense \\ \cline{2-4}
    & Speech Rate 
    & Increased / Decreased 
    & High engagement or urgency / Calm, deliberate, reflective \\ \cline{2-4}
    & Intonation Pattern 
    & Low / Normal / High 
    & Minimal variation (monotone/neutral) / Moderate variation / Dynamic, expressive \\ \hline
\multirow{4}{*}{Non-Verbal Data} 
    & Horizontal Gesture (X-axis Wrist Movements) 
    & High, medium, normal, unilateral/unified motion horizontal 
    & Lateral assertiveness, clarity, or finality (depending on gesture) \\ \cline{2-4}
    & Vertical Gesture (Y-axis Wrist Movements) 
    & High, medium, normal, unilateral/unified motion vertical 
    & Indicates enthusiasm, finality, or focused emphasis \\ \cline{2-4}
    & Hand Configuration 
    & Examples: One hand open \& other closed, both closed, hands together, hands overlapped
    & Reflects receptiveness, assertiveness, introspection, or inactivity \\ \cline{2-4}
    & Expression 
    & Mapped as: Anxiety/Stress (Fear), Calm/Disengaged (Neutral), Aversion (Disgust), Frustration (Sad), Unexpected (Surprise), Positive (Happy), Irritation (Anger) 
    & Represents underlying emotional states \\
\hline
\end{tabular}
\caption{Overview of Sensing, Physiological, and Response Parameters}
\label{tab:sensing_modalities}
\end{table*}

\begin{table*}[ht!]
\small
\renewcommand{\arraystretch}{1}
\setlength{\tabcolsep}{3pt}
\centering
\begin{tabular}{|p{3cm}|p{10cm}|}
\hline
\textbf{Criterion} & \textbf{Concise Scoring Rubric} \\
\hline

\textbf{1. Appropriate Topic} &
\begin{itemize}
    \item \textbf{Advanced (4):} Highly relevant, timely, new info
    \item \textbf{Proficient (3):} Suitable, some new info
    \item \textbf{Basic (2):} Minimal new info, lacks originality
    \item \textbf{Minimal (1):} Trivial/inappropriate for occasion
    \item \textbf{Deficient (0):} No clear topic
\end{itemize}
\\ \hline

\textbf{2. Effective Introduction} &
\begin{itemize}
    \item \textbf{Advanced (4):} Strong opener, clear thesis, credibility established
    \item \textbf{Proficient (3):} Good opener, generally clear thesis, credibility shown
    \item \textbf{Basic (2):} Mundane opener, partial clarity on thesis
    \item \textbf{Minimal (1):} Weak opening, thesis is implied
    \item \textbf{Deficient (0):} No distinct opener or thesis
\end{itemize}
\\ \hline

\textbf{3. Organized Structure} &
\begin{itemize}
    \item \textbf{Advanced (4):} Logical flow, clear transitions, distinct main points
    \item \textbf{Proficient (3):} Mostly logical, uses transitions, main points apparent
    \item \textbf{Basic (2):} Some organization, transitions need improvement
    \item \textbf{Minimal (1):} Disjointed or unclear pattern
    \item \textbf{Deficient (0):} No coherent structure
\end{itemize}
\\ \hline

\textbf{4. Compelling Support} &
\begin{itemize}
    \item \textbf{Advanced (4):} Credible, varied sources, fully cited
    \item \textbf{Proficient (3):} Generally appropriate evidence, mostly cited
    \item \textbf{Basic (2):} Adequate support, citations need clarity
    \item \textbf{Minimal (1):} Insufficient or weak supporting materials
    \item \textbf{Deficient (0):} No credible support or citations
\end{itemize}
\\ \hline

\textbf{5. Strong Conclusion} &
\begin{itemize}
    \item \textbf{Advanced (4):} Memorable closing, ties back to thesis
    \item \textbf{Proficient (3):} Solid wrap-up, brief reference to main idea
    \item \textbf{Basic (2):} Some summary, weak tie-back
    \item \textbf{Minimal (1):} Abrupt or unclear ending
    \item \textbf{Deficient (0):} No recognizable conclusion
\end{itemize}
\\ \hline

\textbf{6. Word Choice} &
\begin{itemize}
    \item \textbf{Advanced (4):} Vivid, precise, bias-free language
    \item \textbf{Proficient (3):} Appropriate language, minimal errors
    \item \textbf{Basic (2):} Some unclear or awkward usage
    \item \textbf{Minimal (1):} Frequent errors, biased terms
    \item \textbf{Deficient (0):} Overly casual, error-ridden
\end{itemize}
\\ \hline

\end{tabular}
\caption{Concise Rubrics for Speech Analysis (Part 1: Criteria 1–6)}
\label{tab:concise_rubrics_part1}
\end{table*}

\begin{table*}[ht!]
\small
\renewcommand{\arraystretch}{1}
\setlength{\tabcolsep}{3pt}
\centering
\begin{tabular}{|p{3cm}|p{10cm}|}
\hline
\textbf{Criterion} & \textbf{Concise Scoring Rubric} \\
\hline

\textbf{7. Audience Adaptation} &
\begin{itemize}
    \item \textbf{Advanced (4):} Clearly tailored, highly relevant examples
    \item \textbf{Proficient (3):} Reasonably adapted, audience interest considered
    \item \textbf{Basic (2):} Minimal tailoring, some relevance
    \item \textbf{Minimal (1):} Weak connection to audience
    \item \textbf{Deficient (0):} No adaptation, ignores audience context
\end{itemize}
\\ \hline

\textbf{8. Persuasive Message} &
\begin{itemize}
    \item \textbf{Advanced (4):} Clear argument, powerful evidence, no fallacies
    \item \textbf{Proficient (3):} Well-reasoned, well-supported, minor gaps
    \item \textbf{Basic (2):} Some logical structure, limited support
    \item \textbf{Minimal (1):} Unclear argument, weak evidence
    \item \textbf{Deficient (0):} No coherent persuasion or reasoning
\end{itemize}
\\ \hline

\textbf{9. Vocal Expression} &
\begin{itemize}
    \item \textbf{Advanced (4):} Excellent modulation, clear enunciation, engaging
    \item \textbf{Proficient (3):} Good variety, mostly clear, few fillers
    \item \textbf{Basic (2):} Some monotony or unclear articulation
    \item \textbf{Minimal (1):} Frequent fillers or volume issues
    \item \textbf{Deficient (0):} Monotone, hard to follow
\end{itemize}
\\ \hline

\textbf{10. Nonverbal Support} &
\begin{itemize}
    \item \textbf{Advanced (4):} Confident posture, purposeful gestures, strong eye contact
    \item \textbf{Proficient (3):} Generally good posture, gestures, and eye contact
    \item \textbf{Basic (2):} Some distracting or stiff nonverbal elements
    \item \textbf{Minimal (1):} Heavily reliant on notes, limited eye contact
    \item \textbf{Deficient (0):} Distracting or absent nonverbal cues
\end{itemize}
\\ \hline

\textbf{11. Dynamic Emphasis} &
\begin{itemize}
    \item \textbf{Advanced (4):} Strategic vocal/physical emphasis of key points
    \item \textbf{Proficient (3):} Generally effective emphasis techniques
    \item \textbf{Basic (2):} Inconsistent emphasis, occasional cues
    \item \textbf{Minimal (1):} Rarely employs intentional emphasis
    \item \textbf{Deficient (0):} No emphasis or misaligned cues
\end{itemize}
\\ \hline

\textbf{12. Emotional Resonance} &
\begin{itemize}
    \item \textbf{Advanced (4):} Verbal and nonverbal alignment enhances authenticity
    \item \textbf{Proficient (3):} Minor mismatches but generally consistent
    \item \textbf{Basic (2):} Some noticeable misalignment at times
    \item \textbf{Minimal (1):} Frequent mismatches undermine authenticity
    \item \textbf{Deficient (0):} No consistent emotional alignment
\end{itemize}
\\ \hline

\end{tabular}
\caption{Concise Rubrics for Speech Analysis (Part 2: Criteria 7–12)}
\label{tab:concise_rubrics_part2}
\end{table*}

\subsection{Data Analysis}

In order to begin the analysis, each participant’s video recording was first converted into text using the OpenWhisper library. This automated transcription process provided a written record of the spoken content, ensuring accuracy and consistency across all samples. To facilitate finer-grained analysis, the transcript was subsequently segmented into 10-second intervals. This segmentation was chosen to capture short, distinct periods of speech and non-verbal actions. For each participant, a separate text file was generated where each 10-second block contained the speech uttered during that specific interval. By structuring the data in this manner, the subsequent steps of feature extraction and prompt construction could be carried out more precisely, as every piece of audio and corresponding text was associated with a defined time frame.

\begin{figure}[ht]
    \centering
    \includegraphics[width=0.5\textwidth]{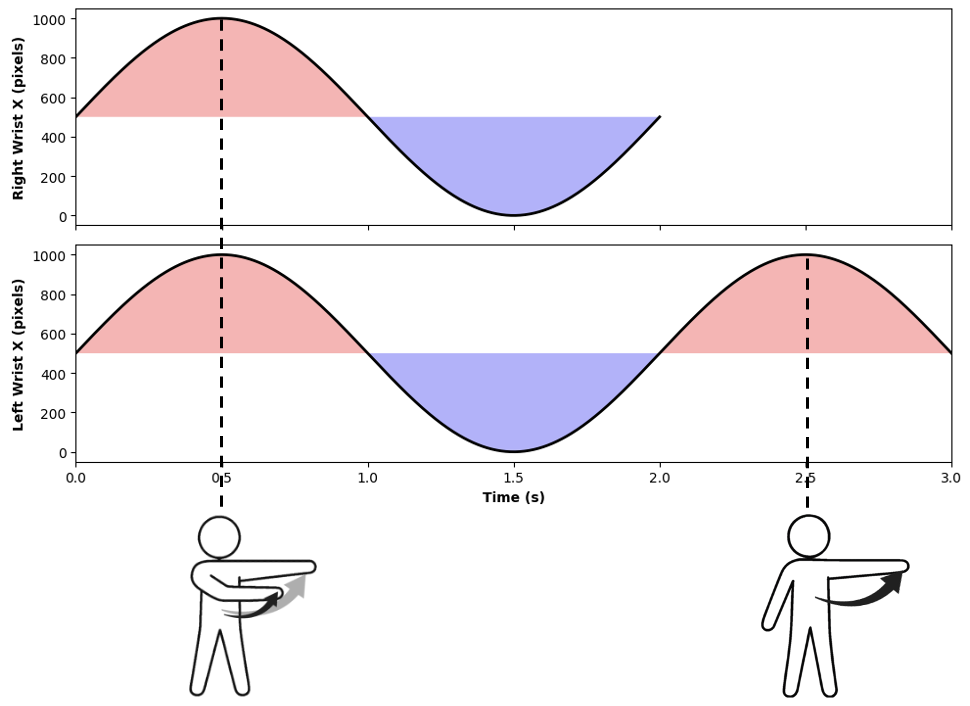}
    \caption{ Wrist Tracking Over Time (X Direction)}
    \label{fig:gesture_capture}
\end{figure}

The next step involved extracting a detailed set of vocal features from each audio snippet. To achieve this, the Librosa library was utilized, as it provides robust methods for analyzing characteristics such as pitch, loudness, speech rate, and intonation patterns. By breaking down the audio into these components, it became possible to quantitatively measure how a speaker’s voice fluctuates over time and how these variations might align with or differ from the textual transcript. For instance, a higher pitch or louder voice might coincide with key moments in the speech, potentially signifying emphasis or heightened emotional states as shown in Table \ref{tab:sensing_modalities}. Similarly, the speech rate and intonation patterns can reveal whether a participant tends to rush through their speech, pause for effect, or maintain a flat vocal tone. These extracted vocal metrics form a critical layer of data that, when paired with the corresponding text segments, enable a more nuanced analysis of each participant’s speaking style and performance across the duration of the recording. Table \ref{tab:sensing_modalities} was constructed based on prior work in multi-modal emotion recognition and nonverbal communication analysis 
\cite{ pavlovic1997visual, eyben2010opensmile, zeng2009survey}.

\begin{figure*}[ht]
    \centering
    \includegraphics[width=0.8\textwidth]{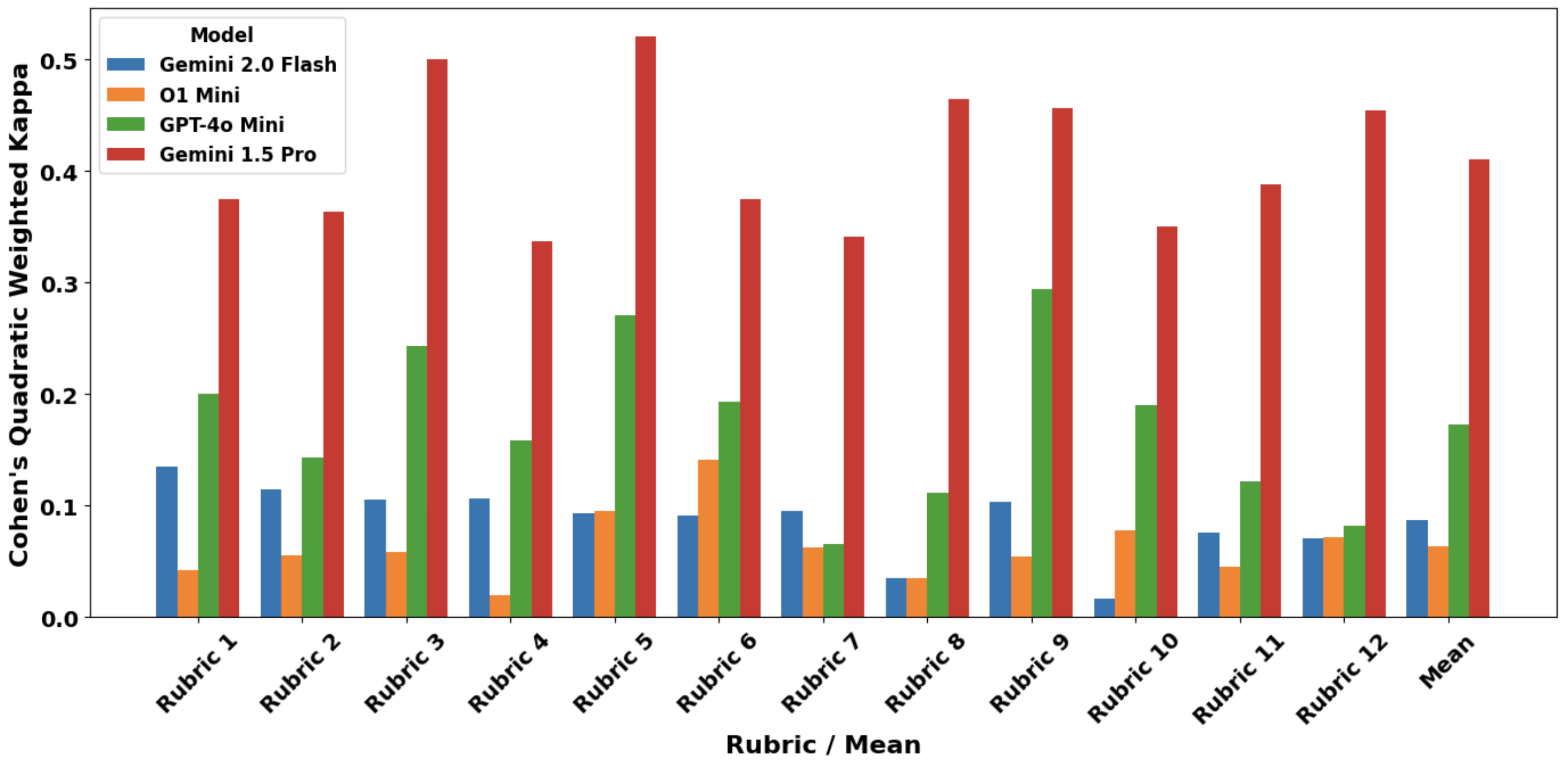}
    \caption{LLM Models vs Human Ground Truth (Kappa Scores)}
    \label{fig:results_example}
\end{figure*}

In addition to analyzing vocal characteristics, an equally vital component was the evaluation of non-verbal indicators. Specifically, facial expressions were identified using the \textit{DeepFace} library, enabling the detection of affective states such as happiness, surprise, or neutrality within each 10-second window. Concurrently, the \textit{Mediapipe} library facilitated the capture of body pose and gestural information, including horizontal and vertical hand movements, wrist coordinates, and overall posture (e.g., whether fists were pointed up or if hands overlapped)  (Table \ref{tab:sensing_modalities}). As illustrated in Figure~\ref{fig:gesture_capture}, the positional data for each hand’s wrist was tracked over time by extracting pixel coordinates from video frames using the Mediapipe framework. In the resulting plots, the y axis represents the horizontal position of the wrist in pixel units (X direction movements). Peaks in these graphs correspond to moments when the wrist reaches its furthest point in one horizontal direction, while valleys indicate the furthest point in the opposite horizontal direction. These peaks and valleys reveal not only the extent of the side-to-side arm movements but also whether both hands are actively involved during a gesture. When the peaks or valleys of the left and right wrist plots occur simultaneously, the motion is classified as unified; otherwise, it is considered unilateral. Furthermore, a similar analysis was performed to capture the Y direction movements---representing the vertical position---to capture upward and downward motions. The combined analysis of both X and Y movements provides a comprehensive view of gesture dynamics, with the cumulative area beneath the curves (peaks or valleys) serving as an indicator of how far the arms extended from the body over the entire speaking duration. A distribution of peak and valley areas was then constructed for each participant’s full speaking duration (1--2 minutes). Based on this distribution, we classified gestures as \emph{normal}, \emph{medium}, or \emph{high} spread, depending on whether their area values fell below the first quartile (Q1), between Q1 and the third quartile (Q3), or above Q3, respectively.

A similar quartile-based method was applied to vocal metrics such as pitch, loudness (RMS), and intonation (rate of pitch change). For each speaker, a “\emph{low}” classification indicated values lying below the 25\textsuperscript{th} percentile for that feature; “\emph{normal}” covered the interquartile range between Q1 and Q3; and “\emph{high}” denoted values exceeding the 75\textsuperscript{th} percentile. By employing these consistent threshold definitions across both the gestural and vocal modalities, the analysis captured a comprehensive picture of each participant’s expressive behavior. Through this multi-layered approach, it was possible to correlate detailed visual cues (e.g., hand movement patterns and posture) with vocal and textual performance, laying a robust foundation for subsequent evaluation of \emph{expressive coherence} and overall presentation quality.

For our analysis, we customized the standard Public Speaking Competence Rubric (PSCR), an 11-item framework for evaluating students’ oral presentation competence. The PSCR addresses key dimensions—content organization, delivery clarity, vocal variation, and non-verbal communication—and provides structured guidance for both formative and summative assessments \citep{schreiber2012pscr}. From the original rubric, we removed the 10th item, which addressed the use of visual aids, given that participants did not employ any such aids in their presentations. Beyond these standard metrics, we introduced two new and more nuanced rubrics: dynamic emphasis---the strategic highlighting of key points through both verbal and non-verbal cues (e.g., a speaker raising their voice and using hand gestures while stressing an important statistic), and emotional resonance---the alignment of emotional content expressed verbally and non-verbally (e.g., sharing a personal story with a soft tone and reflective facial expression to evoke empathy). These rubrics assess how effectively gestures, tone, and expression work together to convey meaning and emotion (Table \ref{tab:concise_rubrics_part2}).

After gathering both vocal and non-verbal features, we combined them with their corresponding 10-second transcript segments into a single, comprehensive file per participant. Each file contained details on precisely what was said (the text from the transcript), the vocal metrics extracted through \textit{Librosa} (pitch, loudness, speech rate, and intonation), and the non-verbal indicators gathered via the \textit{DeepFace} and  \textit{Mediapipe} libraries (facial expressions, body pose, and gestural information) \cite{mcfee2015librosa, serengil2020lightface, lugaresi2019mediapipe}. By consolidating all these elements into a unified record, it became easier to observe how vocal dynamics, facial expressions, and physical gestures related to particular segments of speech. Consequently, these enriched text files served as the foundational input for the subsequent evaluation by LLMs, offering a holistic perspective that bridged the gap between verbal and non-verbal behavior (see Figure \ref{fig:multi-modal_snippet}). For instance, if a speaker exclaimed “I was shocked!” in a high-pitched voice while hand gesturing with open palms, our text file will capture all these three modalities. Such real-world patterns enabled the system to identify expressive coherence, where speech content, vocal tone, and bodily reactions aligned meaningfully.

Building on the rich text file compilations, the next step involved crafting prompts for LLM assessments. Crucially, the prompts were tailored based on whether a particular rubric emphasized textual, vocal, or non-verbal aspects of performance. For rubrics focused exclusively on textual content (1--8) (Table \ref{tab:concise_rubrics_part1}), only the transcript segments were included in the prompt. If a rubric (9 or 10) focused solely on either vocal or non-verbal qualities, the corresponding prompt included the transcript along with the relevant vocal or non-verbal features for each 10-second segment. Meanwhile, rubrics (11 and 12) that addressed both verbal and non-verbal aspects required the full multi-modal dataset, encompassing transcript, vocal metrics, and non-verbal indicators (facial expressions, gestures, and body pose) (Table \ref{tab:concise_rubrics_part2}). The process ensured that each LLM analyzed only the relevant data required to evaluate the specific dimension of public speaking skill defined by the rubric, while maintaining a consistent prompt structure across all participants and time intervals as shown below:

\begin{lstlisting}[language=Python, caption={Common Evaluation Prompt Template}]
"""
You are an expert evaluator. Your task is to judge the data provided below based on the criterion "<Criterion>".

Use the following scoring rubric:
{Rubric}
Use the following definitions to interpret the cues:
{Definition}

Instructions:
1. Analyze the data provided below.
2. Evaluate how well the [persuasive message / vocal expression / non-verbal behavior / dynamic emphasis] is demonstrated based on the above rubric.
3. Return your evaluation as a JSON object with two keys:
   - "score": an integer from 0 to 4.
   - "reason": a brief explanation for the score.

Data:
{RichData}
"""
\end{lstlisting}

In the above template, the variables Criterion and Rubric are dynamically populated with content from (Table \ref{tab:concise_rubrics_part1} and \ref{tab:concise_rubrics_part2}) , iterating through each rubric item individually rather than presenting the entire rubric simultaneously. This approach is designed to yield more effective evaluations. Definition variable provides the model with the understanding of different categories and its physiological response (as shown in Table {\ref{tab:sensing_modalities}}). The variable RichData refers to the comprehensive dataset created for each participant, which includes the transcript, vocal features, and non-verbal information partitioned into 10-second intervals (as shown in Fig \ref{fig:multi-modal_snippet}) for the entire presentation.

\section{Results}
\label{sec:results}

This research makes a distinctive contribution to the pedagogy of public speaking within engineering education. Specifically, it presents three key innovations for leveraging AI in the assessment of public speaking.

\paragraph{A multi-modal LLM-Based Evaluator}
 We propose a novel system, \emph{SapienAI}, which integrates LLMs with speech analysis, computer vision, and sentiment detection. This comprehensive framework simultaneously assesses an individual’s vocal modulation (e.g., pitch, pacing, intonation) and non-verbal expressions (e.g., facial affect, posture, gestural variety), while also measuring what we term \emph{expressive coherence}---the degree of alignment between verbal communication (e.g., clarity, logical flow) and non-verbal cues (e.g., gestures, facial expressions). By capturing how textual meaning, vocal emphasis, and body language work synergistically, SapienAI offers a robust, holistic approach to evaluating public speaking performance. Such an integrated method for multi-modal assessment has not previously been attempted with LLM-based systems and represents a significant advancement in the field.

\begin{table}[ht]
\centering
\begin{tabular}{ll}
\hline
\textbf{Cohen's Kappa} & \textbf{Interpretation} \\
\hline
0                      & No agreement            \\
0.10--0.20             & Slight agreement        \\
0.21--0.40             & Fair agreement          \\
0.41--0.60             & Moderate agreement      \\
0.61--0.80             & Substantial agreement   \\
0.81--0.99             & Near perfect agreement  \\
1                      & Perfect agreement       \\
\hline
\end{tabular}
\caption{Interpretation of Cohen's Kappa values.}
\label{tab:cohen_kappa}
\end{table}

\paragraph{Rigorous Benchmarking with Multiple LLMs}
In order to gauge performance across diverse dimensions of public speaking, we devised a 12-criterion rubric encompassing content organization, vocal expression, and non-verbal aspects. Four distinct state-of-the-art LLMs were tested: \emph{Gemini 1.5 Pro}, \emph{Gemini 2.0 Flash}, \emph{GPT-4o Mini}, and \emph{O1 Mini} (Figure \ref{fig:results_example}). Their evaluations were benchmarked against expert human raters for a cohort of 20 study participants. We used Cohen’s Kappa (weighted) to measure inter-rater reliability. \emph{Gemini 1.5 Pro} achieved the highest overall alignment across all rubrics (mean $\kappa=0.41$), classified as moderate agreement (Table \ref{tab:cohen_kappa}). Moreover, \emph{Gemini 1.5 Pro} surpassed all other models on rubrics integrating multiple modalities; for instance, on Rubric 9 (vocal expression) it attained a Kappa of 0.45, significantly outperforming \emph{GPT-4o Mini} (0.29) and greatly exceeding both \emph{Gemini 2.0 Flash} (0.11) and \emph{O1 Mini} (0.09).
Furthermore, in certain instances, Gemini 1.5 Pro demonstrated superior judgment compared to human graders. For example, the model assigned a score of 1 to a participant (ID 5) whose hands remained stationary and overlapped throughout the presentation, whereas human graders assigned a score of 3, indicating active gesturing. In this instance, our model performed more accurately than the human graders.

Therefore, the moderate agreement for our LLM could encompass two plausible interpretations. On one hand, the discrepancies might stem from the model’s occasional hallucinations, which could be mitigated through techniques such as enhanced post-processing, rigorous model fine-tuning with curated datasets, and the incorporation of cross-validation measures to reconcile automated outputs with human assessments. On the other hand, it is also conceivable that the model is indeed outperforming human graders by effectively tracking and assessing multiple parameters across every frame—a level of detail that is practically unfeasible for human evaluators. This dual interpretation merits further exploration, as it opens the possibility of refining automated evaluation methods while leveraging the model’s capability to deliver nuanced, data-driven insights.
 
\paragraph{Methodological Enhancement}
We introduce the concept of \emph{expressive coherence}, which captures the alignment between verbal communication (e.g., clarity of wording, logical structure) and non-verbal cues (e.g. posture, gestures, facial affect). To operationalize this concept, we extended the standard Public Speaking Competence Rubric (PSCR) by adding two new criteria: \emph{dynamic emphasis} (how verbal and non-verbal signals jointly highlight key points) and \emph{emotional resonance} (the convergence of emotional content through both words and physical demeanor). By incorporating Rubrics 11 and 12 to focus explicitly on expressive coherence, our enhanced rubric demands a more intricate analysis of how text, voice, and body language interrelate. Notably, our Gemini 1.5 Pro based \emph{SapienAI} system performs consistently well across both traditional and newly introduced criteria (Rubric 11 and 12), demonstrating its capacity to handle the holistic demands of public speaking evaluation.

\section{Conclusion}

 Our multi-modal framework, \emph{SapienAI}, combines speech analysis, computer vision, and sentiment detection within a single LLM-based evaluator (Gemini 1.5 Pro), capturing the interplay of verbal articulation and non-verbal cues. Central to this framework is the concept of \emph{expressive coherence}, which elevates the Public Speaking Competence Rubric by adding two criteria—dynamic emphasis and emotional resonance—to quantify how well verbal elements align with gestures, posture, and facial expressions. In benchmarking four distinct LLMs against expert human raters across 20 participants, \emph{Gemini 1.5 Pro} achieved a Cohen’s Kappa of 0.41 (moderate agreement), outperforming all other models and particularly excelling on complex rubrics requiring synergy across text, vocal dynamics, and physical demeanor. Our findings provide initial insights into the capabilities of LLM-based evaluators, highlighting both their potential to offer objective, consistent evaluations and the ability to fully capture the nuanced aspects of public speaking.

Despite the promising results, several limitations remain. Our current feature set does not fully capture the diversity in public speaking styles, which can vary widely across different genders and cultural backgrounds. As a result, the system’s applicability is limited by its inability to account for multiple languages and culturally nuanced behaviors.

Future work should explore the integration of physiological data—such as electrocardiograms, skin temperature, and moisture sensors—to gain richer insights into speaker's stress and emotional states, thereby enhancing the robustness and real-world applicability of automated public speaking evaluation systems. Additionally, addressing the challenges of hallucinations---an issue where LLMs may inadvertently add or misinterpret transcript details---is crucial, as these errors can lead to scores that deviate from a clearly defined rubric. Tackling these problems through improved prompt engineering and verification methods remains an important direction for further research.

\bibliographystyle{IEEEtran}
\bibliography{bibliography.bib}

\end{document}